\pdfoutput=1

\documentclass[11pt]{article}

\usepackage[]{emnlp2021}

\usepackage{times}
\usepackage{latexsym}
\usepackage{graphicx}

\usepackage[T1]{fontenc}

\usepackage[utf8]{inputenc}

\usepackage{microtype}

\title{Ring That Bell: A Corpus and Method for Multimodal Metaphor Detection in Videos} 

\author{Khalid Alnajjar\textsuperscript{1}, Mika Hämäläinen\textsuperscript{1} and Shuo Zhang\textsuperscript{2} \\
  \textsuperscript{1}University of Helsinki, Finland \\
  \textsuperscript{2}Bose Corporation, USA \\
  \texttt{firstname.lastname@\{helsinki.fi or bose.com\}}  \\}

\begin{document}
\maketitle
\begin{abstract}
We present the first openly available multimodal metaphor annotated corpus. The corpus consists of videos including audio and subtitles that have been annotated by experts. Furthermore, we present a method for detecting metaphors in the new dataset based on the textual content of the videos. The method achieves a high F1-score (62\%) for metaphorical labels. We also experiment with other modalities and multimodal methods; however, these methods did not out-perform the text-based model. In our error analysis, we do identify that there are cases where video could help in disambiguating metaphors, however, the visual cues are too subtle for our model to capture. The data is available on Zenodo.
\end{abstract}

\section{Introduction}

Figurative language is a challenging topic for computational modeling as the meaning of a figurative expression is non-compositional and typically very context dependent (see \citealt{roberts1994people}). Metaphor is one of the most important figures of language; it is constantly used in every day language \cite{steen2010metaphor} to draw comparisons or to express something difficult and foreign in more familiar terms. Metaphors can be conventional \cite{traugott1985conventional} and they are often found in idioms, but at the same time metaphors are used to create something new (see \citealt{runoousoppi}). 

Given its ubiquitous presence, understanding metaphors is integral in achieving true natural language understanding (NLU) in the real world. Without their successful interpretation, our models are bound to make mistakes whenever anything is expressed in an indirect or creative fashion. Metaphors are often very contextual and their successful detection and interpretation requires a wide range of contextual cues that would be captured in audio (e.g., prosody) and video (e.g., gestures and actions). Therefore, we believe a multimodal dataset is a great contribution to metaphor research within and outside of the field of NLP.

Two important parts of a metaphor are a tenor and a vehicle (see \citealt{Richards36}). For example, in the metaphor \textit{life is a journey}, \textit{life} is the tenor and \textit{journey} is the vehicle. How metaphors essentially operate is that a vehicle is used to give some of its attributes to the tenor. In the case above, \textit{journeys} are long and full of adventure, which means that these properties are attributed to \textit{life} in an indirect fashion. The meaning of a metaphor is never literal nor compositional, but rather calls for interpretation on the level of pragmatics (see \citealt{rosales2016metaphor}).

Meanwhile, multimodality is becoming increasingly important for many tasks (see \citealt{nlpbt-2020-international,lantern-2020-beyond,mmw-2020-lrec}). We believe the availability of multimodal datasets for a variety of NLP tasks is lacking, and we hope to contribute to the community with our multimodal metaphor dataset.

In this paper, we present the first fully open expert annotated multimodal dataset for metaphor detection\footnote{https://doi.org/10.5281/zenodo.7217991}. In addition, we experiment with uni-modal and multimodal methods for metaphor detection. Our results indicate that the text-based model achieved the best performance. We discuss the results of our experiments and conduct an extensive error analysis to shed light on what was learned successfully by the model and its shortcomings. 

Using CC BY licensed videos in our corpus has been the primary design principle of our data collection so that we can release our corpus without any restrictions in its entirety. This, we believe, is more useful for research purposes than a corpus consisting of short video clips to compile with copyright laws such as the fair use law in the US.

\section{Related Work}

Metaphors have, thus far, been computationally detected using only text. In this section, we describe some of the recent approaches for textual metaphor detection, the corpora used to achieve that and some of the multimodal research conducted on NLP tasks other than metaphor detection. There are several takes on metaphor interpretation \cite{xiao2016meta4meaning,rai2019understanding,bar2020automatic} and generation \cite{hamalainen2018harnessing,terai2019construction,zheng2019love}, but we do not describe them in detail as interpretation is a very different problem.

There are two corpora currently used for metaphor detection, VU Amsterdam (VUA) Metaphor Corpus \cite{steen2010method} and Corpus of Non-Native Written English Annotated for Metaphor \cite{beigman-klebanov-etal-2018-corpus}. Unlike our corpus, both of these datasets contain textual modality only.

For textual metaphor detection, \citet{gao-etal-2018-neural} has used a bi-directional LSTM (long short-term memory) based model with ELMo embeddings. Similarly, \citet{liu-etal-2020-metaphor} have used a bi-LSTM model with BERT and XLNet for the same task. Not unlike the previous approaches, \citet{dankers-etal-2020-neighbourly} has also applied bi-LSTM models comparing ELMo and GloVe embeddings to BERT embeddings with global and hierarchical attention models. Traditional machine learning methods, Logistic Regression, Linear SVC (Support Vector Classification) and Random Forest Classifier, have been used recently with feature engineering to detect metaphors \cite{wan-etal-2020-using}. In DeepMet, proposed by~\citet{su-etal-2020-deepmet}, a siamese neural network have been utilized, where textual RoBERTa~\cite{liu2019roberta} embeddings are computed from the context, the token in question and its part-of-speech and fine-grained part-of-speech. DeepMet was the best performing solution for detecting textual metaphors in the VUA dataset, based on a recent shared task~\cite{leong-etal-2020-report}. 

There are several recent works on multimodal detection of a variety of linguistic phenomena. For example, SVMs (Support Vector Machines) with word embeddings and feature extraction have been used for multimodal sarcasm detection \cite{mustard,alnajjar2021que}. \citet{mittal2020m3er} uses GloVe embeddings, features extracted from audio and facial recognition system output to predict emotion in a multimodal dataset. These multimodal features are fused using a memory fusion network (MFN) \cite{zadeh2018memory}. Similarly, \citet{li2021quantum} detect emotion in a multimodal dataset by modeling the problem from the point of view of the quantum theory. While the field has seen increasing research on multimodal NLP \cite{tsai2019multimodal,mai2020modality,sahu-vechtomova-2021-adaptive}, no data or model has been proposed for multimodal metaphor detection.

\section{Our Metaphor Corpus}

\begin{table*}[!ht]
\centering
\begin{tabular}{|l|}
\hline
\textbf{sentence }                                                                                                                                                          \\ \hline
that you can use to really up your \textless{}v\textgreater{}game\textless{}/v\textgreater{}                                                                       \\ \hline
because while a \textless{}t\textgreater{}quick fix\textless{}/t\textgreater ~can be \textless{}v\textgreater{}appetizing\textless{}/v\textgreater ~and appealing    \\ \hline
\textless{}t r="domain name"\textgreater{}That\textless{}/t\textgreater{}'s \textless{}v\textgreater{}the street address\textless{}/v\textgreater ~for your website \\ \hline
you're ready to \textless{}v\textgreater{}give it a shot\textless{}/v\textgreater{}                                                                                \\ \hline
\end{tabular}
\caption{Example of the annotations for the metaphor detection corpus.}
\label{tab:detection_example}
\end{table*}

In this section, we present our video, audio and textual corpus of manually annotated metaphorical language. Our selection of the video clips includes only CC-BY licensed videos on YouTube that have human authored closed captions in English. The content of the videos presents mainly real people talking, which rules out animations and video game streams. The availability of human authored closed captioning is important as it speeds up our annotation time and provides us with subtitles that are already aligned with video and audio. The CC-BY license was an important selection criterion because it makes it possible for us to release the dataset openly.

We used the filters provided by YouTube to limit our search to videos that were marked as CC-BY and had closed captioning. However, the YouTube filter does not distinguish between automatically generated closed captioning and a human authored one. Fortunately, it is relatively easy to tell these two apart from each other. Automated closed captioning tends to appear one word at a time, whereas human authored closed captioning is visualized more like traditional subtitles. These criteria greatly reduced the number of eligible videos to include in our corpus. Apart from these criteria, we also filtered videos with sensitive and offensive languages. No further restrictions have been explicitly placed on the genres or types of videos, as we do not want to introduce biases for which types of contents are more likely to contain metaphors. Therefore, the availability of the metaphors naturally occurring in the corpus is the result of the ubiquity of the metaphor in everyday language use. All Youtube queries were conducted in incognito mode to avoid biased YouTube suggestions based on our viewing habits.

\begin{figure}[h]
  \centering
  \includegraphics[width=\linewidth]{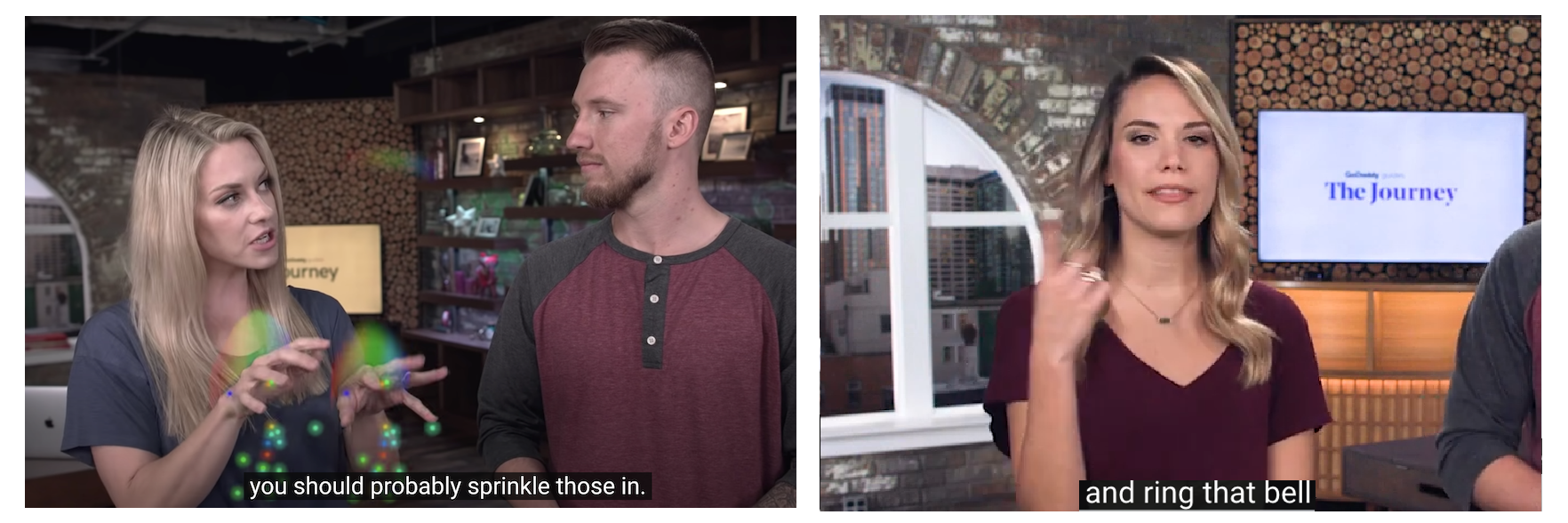}
  \caption{Metaphors made visible in the video through gestures.}
  \label{example_archer}
\end{figure}

Figure \ref{example_archer} shows real examples from our corpus where video can be useful in detecting metaphors. On the left, the woman wearing a gray shirt is talking about \textit{sprinkling keywords} and showing a sprinkling gesture. On the right, the woman wearing the wine red shirt says \textit{ring that bell} and shows a bell ringing gesture.

Our corpus consists of 27 YouTube videos with a total duration of 3 hours, 53 minutes and 47 seconds of video. For comparison, a recently released multimodal dataset for sarcasm detection \cite{mustard} has the duration of 3 hours, 40 minutes and 47 seconds.  The videos belong mostly to a start-up domain and many of them deal with issues of online visibility for a start-up company. This domain was a consequence of our selection criteria for videos. It turns out that YouTube has plenty of high-quality human close-captioned videos released under the CC-BY license that relate to this particular domain.

Our corpus provides linguistics researchers with the ability to study the use of metaphor in a multimodal setting, something that has gained attention in their field of science as well~\cite{MullerCienki}. This can, indeed, foster a wider interdisciplinary collaboration leading to a deeper understanding of the phenomenon.

\subsection{Annotation}

Two expert annotators went through the video files and annotated metaphors by surrounding them with \textit{v} tags for vehicles and \textit{t} tags for tenors. The use of experts is motivated by the fact that previous research has found that non-expert annotators struggle with metaphors \cite{hamalainen-alnajjar-2019-lets}. 

The annotators followed a simple procedure in annotating the data:

\begin{itemize}
  \item Is the meaning literal?
  \item If the meaning of the word is abstract, is it a dictionary meaning?
  \item Does the potential metaphor express pragmatic insincerity?
  \item If the answer to all of the questions is no, annotate it as a metaphor.
\end{itemize}

In other words, if the meaning of a word or a phrase is not literal, it is annotated as a metaphor. However, just the mere fact of a word being used in an abstract way is not enough to mark it as metaphorical. For example, in the sentence \textit{it is tied to revenue}, ``tied'' is not tagged as a metaphor just because it is used in a more abstract sense than the typical concrete sense of tying one's shoes, for example. If the abstract meaning of a word appears in a dictionary, the word is not considered metaphorical. However, conventional metaphors that consist of multiple words, and are thus idioms, are tagged as metaphors. We do not make a distinction between metaphors and similes.

Pragmatic insincerity (see \citealt{grice1975logic}) is a phenomenon related to sarcasm as one of its preconditions (see \citealt{kumon1995another}). There is a certain overlap between metaphors and sarcastic expressions in the sense that both use words in their non-literal meaning. In order to ensure that we do not mix these two notions with each other, it is important to avoid annotating pragmatically insincere expressions as metaphorical.

Table \ref{tab:detection_example} shows an example of annotations. The annotations were done directly in the subtitles The utterances are time stamped and aligned with the video. In the table, tenors are indicated with \textit{$<$t$>$} and vehicles with \textit{$<$v$>$}. For deictic tenors, an \textit{r} attribute is provided to resolve the deixis by indicating the actual tenor that has appeared earlier in the conversation. In the examples, \textit{game} is used metaphorically to talk about marketing, \textit{quick fix} is called \textit{appetizing} as though it was something edible and \textit{domain name} is contrasted to a physical \textit{street address} by direct comparison. \textit{Give it a shot} is a conventional metaphor.

All in all, after multiple annotation iterations, the dataset consists of 304 vehicles and 67 tenors. This totals to 371 metaphorical expressions. They vary in length: the shortest tenor is one word, such as \textit{it}, while the longest tenor is several words \textit{the discovery of those five noble gases to illuminate like that}. The same goes for vehicles where their length varies form one word such as \textit{dive} to multiple words: \textit{the history of the internet itself}. On a token level, we have 672 vehicle tokens and 113 tenor tokens, so altogether 785 metaphorical tokens. 

In total, 6\% of the expressions in the corpus are metaphorical. While this percentage might appear low, it is natural and more representative of the real usage of metaphors in typical conversations which makes this corpus suitable for building metaphor detection models applicable for real-world scenarios.

Around 55\% of the vehicles are conventional metaphors and 45\% are novel metaphors. However, it is fairly common that same words appear in the corpus in a metaphorical and non-metaphorical sense. In our corpus, there are two videos that deal with actual cooking, in which many food-related metaphors appear non-metaphorically, such as \textit{sprinkle those in}, said metaphorically about keywords and \textit{a little sprinkle}, said non-metaphorically about sugar. Another example is the use of \textit{house} non-metaphorically as in \textit{come pick it up at my house} and metaphorically as in \textit{think of hosting as your house}, where a metaphorical connection is drawn between \textit{hosting} and a \textit{house}.

\subsection{Data preparation}

As YouTube serves files in several different formats such as \textit{webm}, \textit{mkv} and \textit{mp4} the first step is to use FFmpeg\footnote{https://ffmpeg.org/} to convert all videos into mp4 format. We also use the same tool to clip the video files into sentence-length clips based on the time stamps in the subtitles and extract their audio into wav files. This process yielded 6,565 video and audio clips that are aligned with text.

We split the datset randomly so that 70\% of sentences that contain metaphors and 70\% of sentences that don't contain any metaphors are used for training, 15 \% of both types of sentences for validation and 15\% of both for testing. This way we ensure that both metaphorical and non-metaphorical sentences are divided proportionally with the same ratios. These splits are used for all the models.

\section{Metaphor Detection}
We experiment with uni- and multi-modal models for metaphor detection. In this section, we describe the preprocessing steps applied and the experimental setups conducted.

\subsection{Preprocessing}
For each modality, we make use of the latest advances in neural network models to capture important features that have achieved state-of-the-art results in various NLP tasks. As metaphor detection has been conducted solely based on text, we follow the DeepMet approach by~\citet{su-etal-2020-deepmet} and process the entire textual content using spaCy~\cite{spacy} to tokenize it and acquire Universal Dependencies style syntactic trees \cite{nivre2020universal} and Penn Treebank parts-of-speech tags \cite{santorini1990part}. 
Similarly to the original approach, all of our textual models predict metaphors at the token level given the context surrounding it and its POS tags as input.

We resample the audio to 16kHz. Audio features are extracted using \textit{Wav2Vec2FeatureExtractor} provided by the Transoformers Python library~\cite{wolf-etal-2020-transformers}. 

Video features are obtained by taking equally-distributed 16 frames from a clip and then resize them into 128x171, followed by normalization and center cropping to 112x112.

\subsection{Textual model}
We train two text-only models, both follow the architecture and approach of DeepMet where we obtain textual embeddings using RoBERTa~\cite{liu2019roberta} and feed them into two transformer encoding layers which are then combined by applying global average pooling and concatenation. A dense fully-connected layer takes in the combined output of both encoders and predicts whether the token is metaphorical (c.f.,~\citealt{su-etal-2020-deepmet} for more details).

In our first textual model, we train the model using our corpus, whereas in the second one we train it using VUA corpus (with a learning rate of 0.00001, akin the original paper) and later fine-tune it using our corpus.

\subsection{Audio model}
We extend and fine-tune Facebook's pretrained multilingual XLSR-Wav2Vec2 large model~\cite{baevski2020wav2vec}. The model is trained on Multilingual LibriSpeech \cite{Pratap2020MLSAL}, CommonVoice \cite{ardila-etal-2020-common} and Babel \cite{roach1996babel} for speech recognition. 
We employ this model to encode speech into vector representations from raw audio.

We replace the classification layer of the original model with a dense fully-connected layer that produces two outputs, one for each label. Unlike the textual model, here we classify whether the entire spoken expression contains a metaphor or not (i.e., not on a word level).

\subsection{Video model}
For our video unimodal model, we incorporate a pretrained model for human action detection. The model is based on the 18 layer deep R(2+1)D network~\cite{8578773} and it is trained on the Kinetics-400~\cite{46330} dataset. The intuition behind using this model is that it was able to detect actions (e.g., playing organ), gestures (e.g., pointing) and movements (e.g., waving). Realizing such information is crucial in understanding the context, and would provide further cues for detecting metaphors.

Similar to the audio model, we substitute the original classification layer with a fully connected layer and fine-tune the pretrained model to predict whether a scene is metaphorical or not.

\subsection{Multimodal metaphor detection}

We test out three multimodal metaphor detection models; 1) text and audio, 2) text and video and 3) text, audio and video. The textual model is the fine-tuned model using the VUA corpus and our textual corpus. 

In all of the models, the final classification layer of their sub-models are removed. Unimodal models are combined by concatenating the weights of their last layer, which are then fed to a classification layer.

\subsection{Common configuration}
All of the models described above share common configurations, unless we explicitly indicate otherwise. Prior to the last classification layer of all of our mono- and multimodal models, we introduce a dropout layer~\cite{JMLR:v15:srivastava14a} (with a probability of 20\%) to accelerate training, and reduce internal covariate shift and overfitting. 

We use the cross entropy loss function along with Adam optimizer~\cite{kingma2014adam,loshchilov2019decoupled} to update the weights and train the models. All the fine-tuned models are trained with a learning rate of 0.0001 and for 3 full epochs.

\section{Results}
In this section, we follow the evaluation metrics commonly used for the metaphor detection task by reporting the precision, recall and F1 scores for the metaphorical label.

Regarding the textual models, we report three sets of results, which are for the models trained on: 1) VUA corpus, 2) our corpus and 3) both the VUA and our corpus. All the models predict metaphoricity on the token level. To ensure that our implementation of the DeepMet approach is correct, we tested the first model on the VUA test dataset of the metaphor detection shared task and achieved an F1-score of 0.68 and 0.73 on all POS and verb subsets of the data, respectively. These results are relatively close to the results reported by the authors. 

Table~\ref{tab:text-results} shows the classification results of all three models on the test set. The test set contained 90 metaphorical tokens and 6,961 non-metaphorical tokens. The results indicate that the textual model trained solely on the VUA dataset performed poorly on our test set. In comparison, training the model using our metaphor corpus only resulted in a great increase of correct predictions. Nonetheless, combining both corpora by fine-tuning the first model with our corpus produced the winning model, which managed to spot 76\% of the metaphorical tokens correctly. 

We believe that the huge differences between the first and second textual models, despite the larger size of VUA's training dataset, are due to the differences in domains. The VUA corpus contains academic texts, conversation, fiction, and news texts, whereas our corpus is dominated by conversations on the web and start-ups. It is evident that by exposing the model to general domains (i.e., VUA's corpus) and, thereafter, concentrating it on the start-up domain, the model was able to identify the highest number of metaphorical usages.

\begin{table}[]
\centering
\begin{tabular}{|l|c|c|c|}
\hline
\textbf{Trained on}  & \textbf{Precision} & \textbf{Recall} & \textbf{F1-score} \\ \hline
VUA & 0.04      & 0.33   & 0.07     \\ \hline
Ours & 0.38      & 0.63   & 0.47     \\ \hline
VUA + Ours & \textbf{0.53}      & \textbf{0.76}   & \textbf{0.62}     \\ \hline
\end{tabular}
\caption{Classification results of the textual monomodal models on the test set of our corpus, for the metaphorical label.}
\label{tab:text-results}
\end{table}

Results from the other models (unimodal or multimodal) that involving audio and video showed that adding these modalities actually did not help improving the model - rather, they are detrimental to the model performance on metaphor detection. We extend two possible explanations for this failure. First, it is possible that because the visual and audio cues of metaphor are subtle, these models failed to learn from such a small amount of annotated data. 

Second, it is unclear that the specific models we are using for audio and video modalities encode the information relevant for the metaphor detection task. For instance, whereas it is impossible to completely disentangle what exactly the Wav2Vec model is encoding, we can conjecture that it encodes information about phoneme identity considering it is optimized for the speech recognition task. Therefore, it may not be entirely surprising that the Wav2Vec encoding is not useful for the metaphor detection task because it is adding redundant or irrelevant information to the model. It is our future work (or the future work for the community who utilizes this dataset) to refine our understanding of the multimodal encoding for the metaphor detection task (for instance, employing a model that more directly encodes information about speech prosody from the audio).




\subsection{Error analysis}

When looking at the results of the text only model, we can see that the model identifies metaphors correctly as metaphors more often than not. There are some metaphorical tokens in metaphors consisting of multiple words that get classified wrong, for example, in \textit{You could think of hosting as your house}, the tenor \textit{hosting} and the determinant \textit{your} of the metaphorical word \textit{house} are not identified as metaphorical, while \textit{house} is correctly identified. Another example is the conventional metaphor \textit{toot their own horn}, where all other words except for \textit{own} are correctly identified as metaphorical. 

There are also a fewer number of cases where all words get identified wrongly as non-metaphorical, for example, the model did not predict any metaphorical tokens in \textit{It's where you live}, while in reality \textit{it} is the tenor and \textit{where you live} is the vehicle. Also, individual tenors where the vehicle comes later get often not recognized such as in \textit{Yes, malware you could think of like}, where \textit{malware} is the tenor for a vehicle that appears later in the dialog.

When the tenor and the vehicle co-exist nearby, the model can get all metaphorical tokens right such as in \textit{It's kinda like real estate right?} where both the tenor \textit{it} and the vehicle \textit{real estate} are correctly identified. Also many tenorless expressions are fully recognized correctly as metaphorical, such as \textit{Spreadin' the love}.

There were plenty of cases (61) where the model predicted a metaphor tag for a token while there was no metaphor. Curiously, prepositions were often tagged metaphorical, such as \textit{to} in \textit{ring that bell to see these episodes first}. The actual metaphorical part \textit{ring that bell} ends before the preposition \textit{to} that has a non-metaphorical meaning \textit{in order to}. 

We can also see that the model was indeed fooled by cooking terms that were used both metaphorically and non-metaphorically. In \textit{Yeah a little sprinkle}, both \textit{a} and \textit{sprinkle} were classified as metaphors, while the context was about sprinkling sugar. Another similar case was \textit{there's five noble gases that illuminate}, where \textit{noble gases} and \textit{illuminate} were erroneously classified to be metaphorical. This was clearly due to the tenor in the corpus: \textit{the discovery of those five noble gases to illuminate like that} that contained similar words. It is evident that the model relies on word similarities more than reaching to a higher pragmatic representation of the phenomenon, however, this is not an unexpected behavior from a machine learning model.

There are also cases where the model detects a metaphor, that could theoretically be a metaphor, but is not because of the way it was used in the corpus. For example, the model predicts \textit{Give it a go} as metaphorical in the expression \textit{button, "Give it a go."}, where people are talking about a button with a particular text rather than using the expression metaphorically. Another such an example is \textit{flying} in \textit{(money flying)}. Such an expression might be used metaphorically, but in this case this was a note for the hearing impaired as money was actually flying on the video.

\section{Discussion and Conclusions}

\begin{figure}[h]
  \centering
  \includegraphics[width=\linewidth]{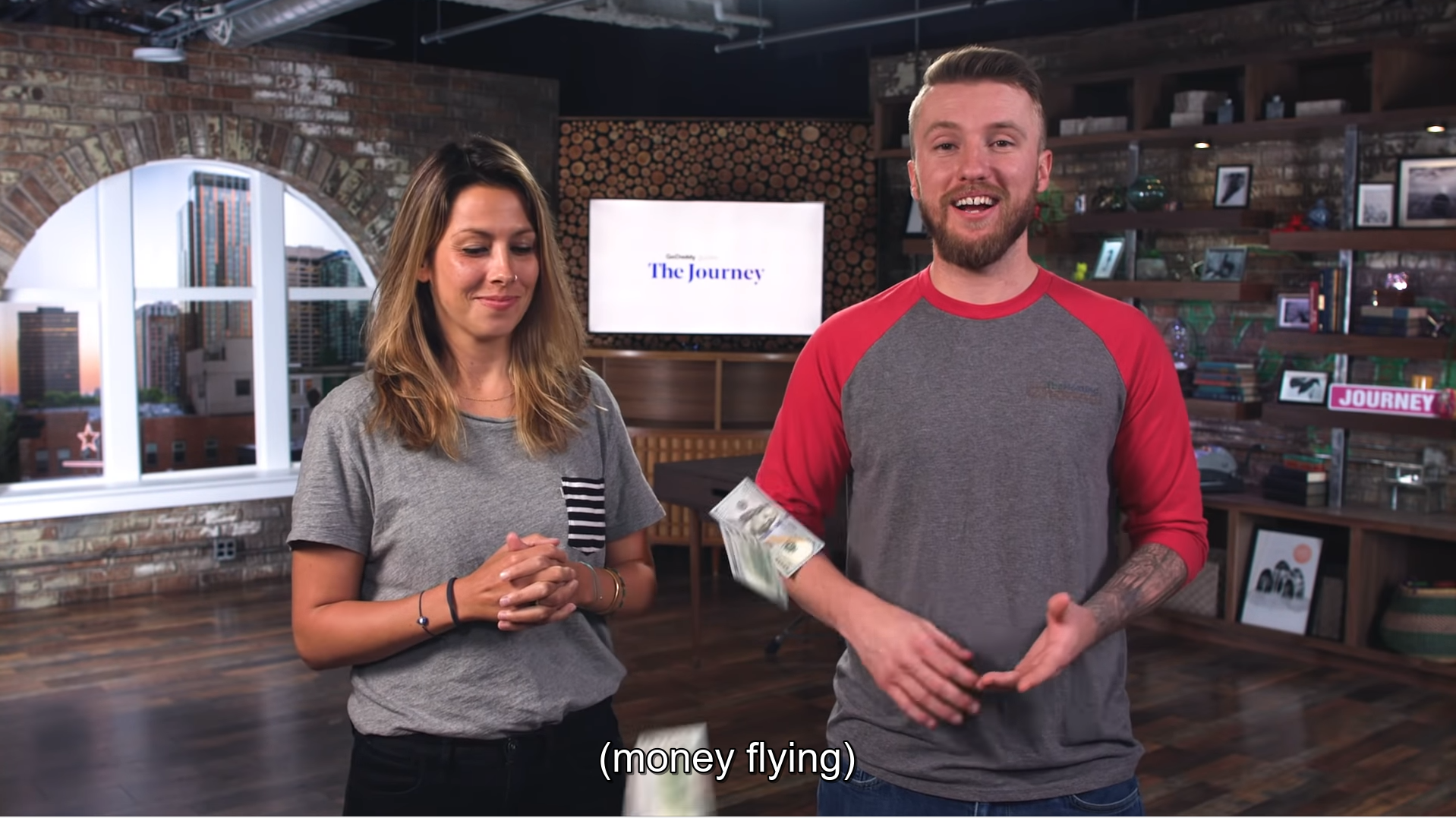}
  \caption{Money actually flying on the video.}
  \label{money}
\end{figure}

In this work, we have only focused on metaphor as a strictly linguistic phenomenon and we have built a multimodal dataset where these linguistic metaphors have been tagged in terms of tenors and vehicles. However, it is apparent that metaphor is a phenomenon that occurs on a higher level of our cognitive capacities than mere language. There are several cases in our corpus, where we can evidence the existence of a metaphor but it is never expressed verbally. For example in Figure \ref{money}, \textit{money flying} cannot be a metaphor when inspected purely from the point of view of language and its relation to the video when money is actually flying in the scene. However, it is a metaphor on a higher level in the sense that the entire scene where money was flying was to indicate someone becoming rich. In other words, stating a fact that is happening is not metaphorical if the fact is literally taking place, however the fact itself might be metaphorical.

\begin{figure}[h]
  \centering
  \includegraphics[width=\linewidth]{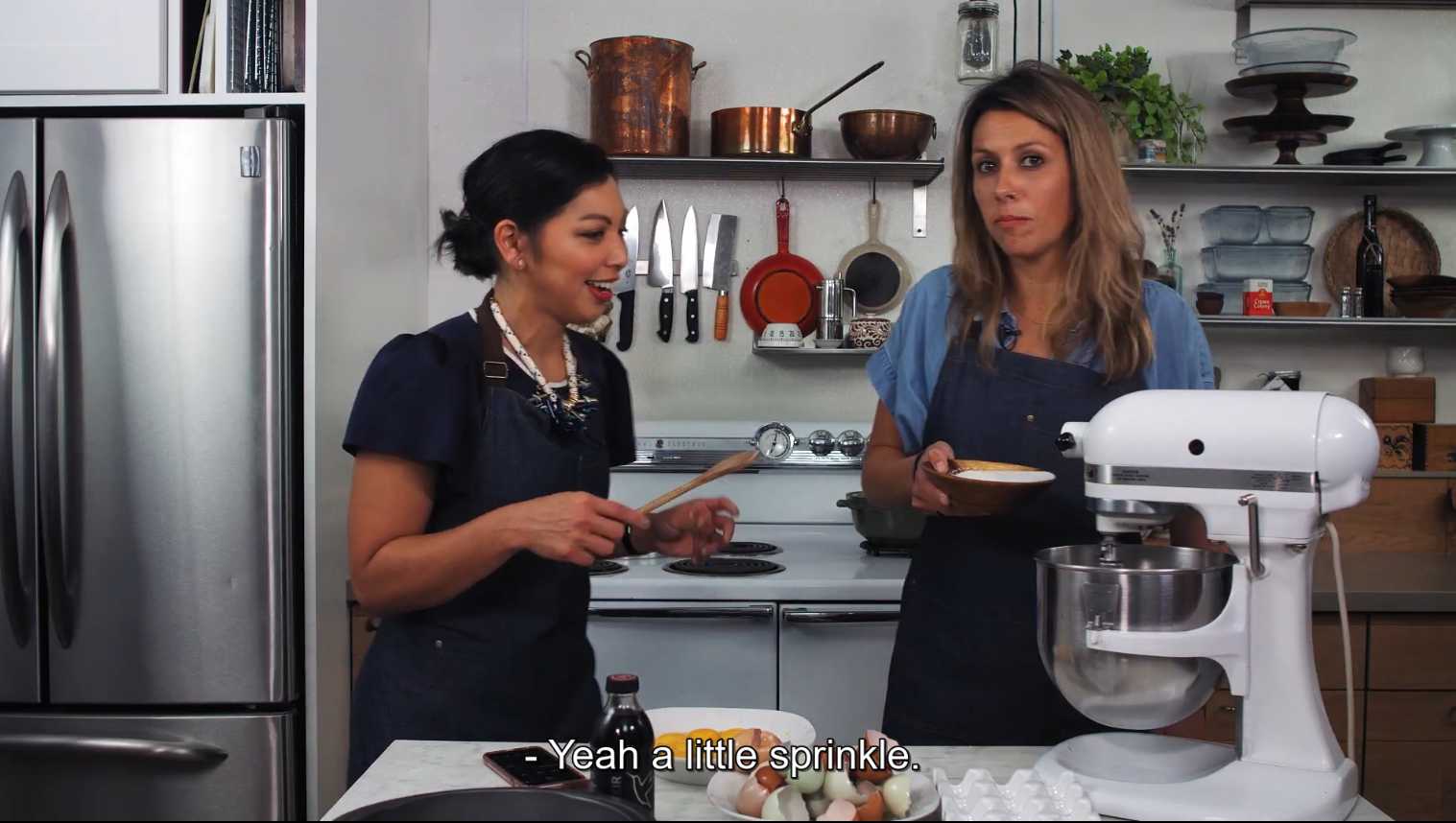}
  \caption{\textit{Sprinkling} used in a kitchen in reference to sugar.}
  \label{sugar}
\end{figure}

At the same time, as evidenced by our error analysis, there are certainly cases where video modality could help in disambiguating whether something is said metaphorically or not. For instance, talking about \textit{sprinkling} in a kitchen environment (see Figure \ref{sugar}) is a very strong sign that the word is potentially non-metaphorical. Integrating these weak cues into a multimodal system is, however, not an easy task given that the current methods for video processing are limited in their coverage.

Therefore, in the future, it would be useful to annotate metaphors also in the other modalities. Money flying can be a visual metaphor, and so can a sound effect, and they can exist independently from each other in different modalities. Perhaps the reason why our multimodal attempts failed was that metaphor can be independent of the other modalities. Producing such a dataset where these modal specific metaphors are also annotated for video and audio is definitely a huge undertaking that requires research in its own right.

It is clear that our model can detect metaphors correctly, but also the mistakes it makes highlight that despite using a large RoBERTa model, the meaning representation the model has cannot reach to such a nuanced level as to confidently detect metaphors. Metaphor is a figurative device that cannot be explained by semantics, but rather requires pragmatic inspection. It is not clear based on our research and other contemporary approaches whether the current word or sentence embedding models are sufficient to navigate in the depths of pragmatics and subjective interpretation in any other way than learning some irrelevant co-occurring phenomena from a biased corpus. At the same time there is no such thing as an unbiased corpus, either, given that bias (and mostly heuristics causing it) is a fundamental part of our cognition as human beings.

In this paper, we have presented a new open and multimodal dataset for metaphor detection. Because we have focused strictly on CC-BY licensed videos, we can make the entire dataset available on Zenodo. In our current work, we have not taken the context widely into account when predicting metaphoricity, but rather resorted to a very local context. The fact that the videos can be published in full length makes it possible for any future work to explore different ways of including contextual cues freely. 

\section*{Acknowledgments}
This work was partially financed by the Society of Swedish Literature in Finland with funding from Enhancing Conversational AI with Computational Creativity, and by the Ella and Georg Ehrnrooth Foundation for Modelling Conversational Artificial Intelligence with Intent and Creativity.

\bibliography{anthology,custom}
\bibliographystyle{acl_natbib}

\end{document}